\documentclass[a4paper]{article}
\usepackage{INTERSPEECH2018}
\usepackage{multirow}
\usepackage{color}
\usepackage{lineno}
\usepackage{cite}

\title{Twin Regularization for online speech recognition}
\name{Mirco Ravanelli, Dmitriy Serdyuk, Yoshua Bengio}
\address{
  Montreal Institute for Learning Algorithms (MILA) \\
  Universit\'e de Montr\'eal, Canada}
\email{mirco.ravanelli@gmail.com, serdyuk@iro.umontreal.ca}

\begin{document}

\maketitle
\begin{abstract}
Online speech recognition is crucial for developing natural human-machine interfaces. This modality, however, is significantly more challenging than off-line ASR, since real-time/low-latency constraints inevitably hinder the use of future information, that is known to be very helpful to perform robust predictions. 

A popular solution to mitigate this issue consists of feeding neural acoustic models with context windows that gather some future frames. This introduces a latency which depends on the number of employed look-ahead features. 

This paper explores a different approach, based on estimating the future rather than waiting for it. Our technique  encourages the hidden representations of a unidirectional recurrent network to embed some useful information about the future. Inspired by a recently proposed technique called Twin Networks, we add a regularization term that forces forward hidden states to be as close as possible to cotemporal backward ones, computed by a  ``twin" neural network running backwards in time. 

The experiments, conducted on a number of datasets, recurrent architectures, input features, and acoustic conditions, have shown the effectiveness of this approach. One important advantage is that our method does not introduce any additional computation at test time if compared to standard unidirectional recurrent networks.

\end{abstract}


\section{Introduction}

\emph{Automatic speech recognition} (ASR) has made great strides in recent years \cite{lideng}. Deep learning, in particular, has contributed to radical transformations in the field, allowing current technology to reach unprecedented performance levels \cite{Goodfellow-et-al-2016-Book,dahl2012context}.  Despite the impressive achievements of the last years, many open challenges remain in the field \cite{watanabe_book}. 

One important issue is the performance drop observed when going from off-line to online speech recognition. The latter recognition modality is significantly more challenging than off-line ASR, due to the real-time/low-latency constraints which inevitably arise. To provide a speech transcription with low-latency, the speech decoding must start while acquiring the signal itself, forcing the acoustic model to perform predictions mostly  based on current and past information. 
Future information plays an important role to perform robust predictions, due to both  phoneme co-articulations and  linguistic dependencies \cite{graves}.

Despite its complexity, online speech recognition is a key component towards a more natural human-machine interaction, and extensive effort has been devoted in the last decade to improve this technology. Past online recognizers were based on the GMM-HMM framework \cite{acero_book}, while current solutions rely on deep learning \cite{Goodfellow-et-al-2016-Book}. In particular, the use of feed-forward Deep Neural Networks (DNNs), including both fully-connected and convolutional architectures, has been largely investigated in the literature \cite{acw1,bacchiani_online}, especially in the context of online ASR performed on small-footprint devices \cite{small2,small3,small4,online2,lu_smallfootprint}. Attempts have also been made to develop robust online speech recognizers based on RNNs, exploiting both the traditional RNN-HMM framework \cite{unirnn_online,peddinti_online,bidir_online1,online_zeyer,bidir_online_chen,bidir_mohamed} and, more recently, end-to-end ASR technology \cite{online_e2e,baidu}. 

A common aspect of past approaches is that they often employ asymmetric context windows \cite{RAVANELLI2018,tdnn2,ravanelli15}, that embed more past than future information.  Despite their effectiveness, context windows inevitably introduce a trade-off between latency (that depends on the number of look-ahead frames) and recognition accuracy. Moreover window-based approaches focus on short-term future dependencies only, while long-term information cannot be used without incurring an unacceptable latency. 

In contrast to past work, this paper attempts to predict the future rather than waiting for it. 
Our technique encourages the hidden representations of a unidirectional recurrent network to embed some relevant features about the future, providing useful information on the upcoming phonetic and linguistic dependencies. Inspired by a recently proposed technique called \textit{Twin Networks} \cite{twin_ref}, we add a regularization term that forces forward hidden states to be as close as possible to cotemporal backward ones, computed by a ‘‘twin” neural network running backwards in time. The twin backward network is employed at training time, when online constraints do not arise. At test time only the forward states are computed, leading to a model that (ideally) does not introduces any latency and does not add any computation compared to standard unidirectional recurrent networks. 

The experiments, conducted on several datasets, recurrent architectures, input features, and acoustic conditions, have shown the effectiveness of our approach. 
To summarize, the contribution of this paper is two-fold.
Firstly, we design a novel method for online ASR that predicts future states of the model and demonstrate that it is well motivated. Secondly, we evaluate the proposed method under a variety of experimental conditions, showing its effectiveness against strong baselines. 

The rest of the paper is organized as follows. Twin regularization for online ASR and the related work are outlined in Sec.~\ref{sec:twin}. The experimental setup and the results are presented and discussed in Sec.~\ref{sec:setup} and Sec.~\ref{sec:exp} respectively. Finally, Sec.~\ref{sec:conc} draws our conclusions.


\section{Twin Regularization for Online ASR} \label{sec:twin}

\begin{figure}[t!]
\centering
  \includegraphics[scale=0.45,trim={0cm 5.5cm 20cm 0},clip]{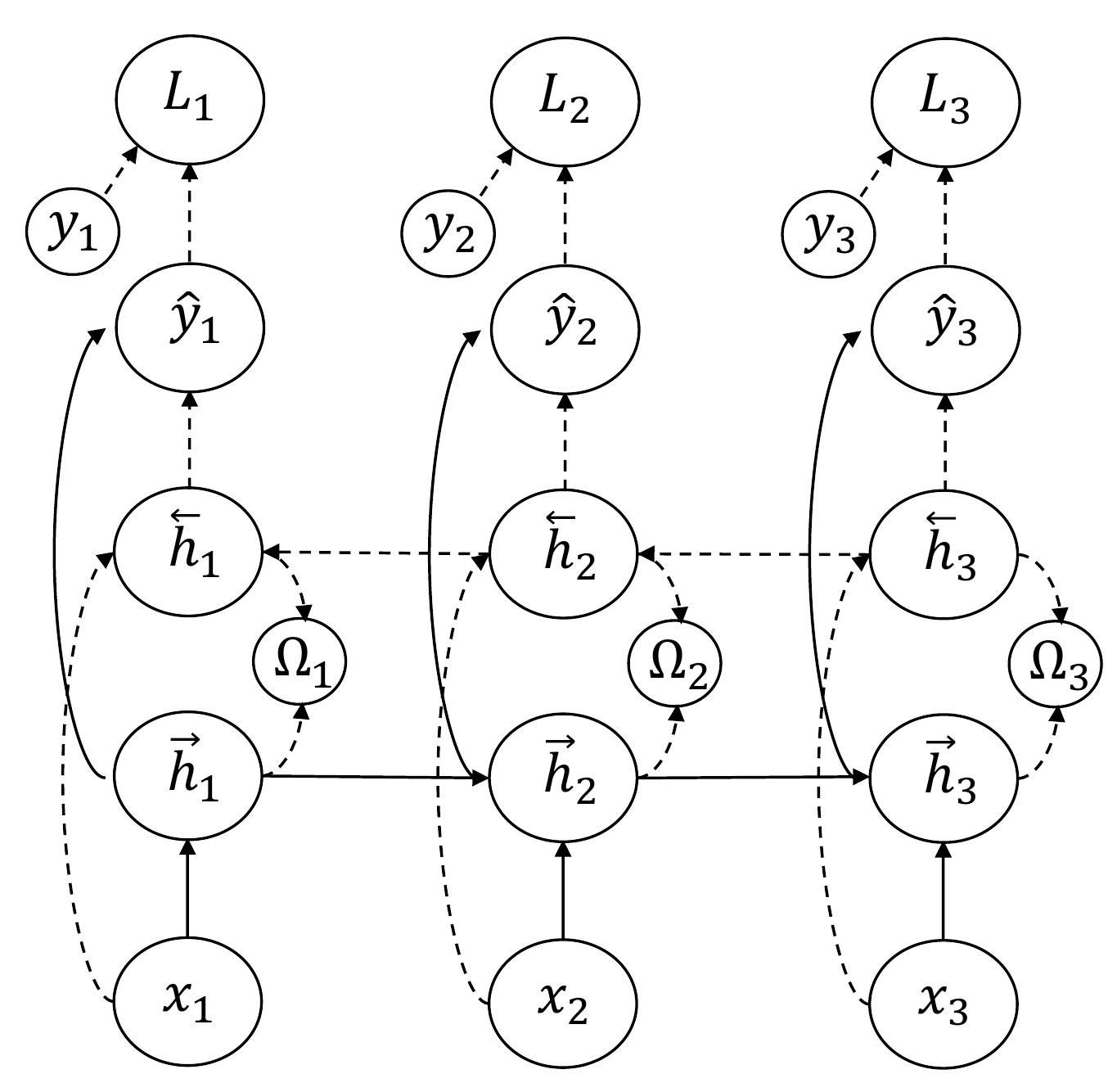}
\caption{An example of a recurrent acoustic model extended with a Twin Network. The regularization term $\Omega$ encourages forward and backward hidden states to be as close as possible. Dashed lines refer to computations done at training time only.}\label{fig:twin}
\end{figure}

In the context of hybrid RNN-HMM speech recognition, the neural network processes the input speech sequence \mbox{$X=\{x_{1},...,x_{t},..,x_{N}\}$} and computes, at each time step $t$, a hidden state in the following way:
\begin{equation}
\label{eq:forward}
\overrightarrow{h_{t}}=f_{\overrightarrow{\theta}}(x_{t},\overrightarrow{h}_{t-1}),
\end{equation}
where $f_{\overrightarrow{\theta}}$ is a function that depends on trainable parameters $\overrightarrow{\theta}$ (such as an LSTM or a GRU cell), and $\rightarrow$ highlights that the network scans the sequence in the forward direction. The forward states summarize the information about current and past elements of the sequence.
A linear transformation, followed by a softmax classifier, is then employed to perform predictions $\widehat{Y}=\{\widehat{y}_{1},...,\widehat{y}_{t},..,\widehat{y}_{N}\}$ over a set of phone states. 

These posterior probabilities, after being normalized by their prior, feed a HMM-based decoder that integrates acoustic and linguistic information into a search graph and estimates the sequence of words uttered by the speaker. The decoding step is normally very computationally demanding, especially for large vocabulary speech recognition. For the unidirectional RNN models described above, however, the output $\widehat{y}_{t}$ at each time step $t$ can be computed without waiting for the full speech utterance to finish. The decoding step can thus start while acquiring the speech signal from the user. 

The RNN model is trained to optimize the \emph{negative log-likelihood} (NLL) cost function:
\begin{equation*}
L(X,Y;\overrightarrow{\theta})=-\frac{1}{N}\sum_{t=1}^{N}{\log P_{\overrightarrow{\theta}}\left(y_{t}| \{x_0,..,x_t\}\right)},
\end{equation*}
where $Y=\{y_{1},...,y_{t},..,y_{N}\}$  is the sequence of targeted phone labels and $P_{\overrightarrow{\theta}}(y_{t}| \{x_0,..,x_t\})$ is the output probability estimated by the neural network (that is  given by reading the entry for $y_t$ from the RNN output vector $\widehat{y}_t$).

When possible, it is very convenient to also process the speech sequence in the reverse time order, and compute backward states similarly to Eq.~\ref{eq:forward}:
\begin{equation*}
\overleftarrow{h_{t}}=f_{\overleftarrow{\theta}}(x_{t},\overleftarrow{h}_{t+1}).
\end{equation*}

The backward states summarize the information about current and future elements of the sequence.
In standard Bidirectional RNNs~\cite{bidir_schuster}, forward and backward hidden states are combined to perform predictions based on the whole speech sequence. This leads to a substantial performance improvement in ASR~\cite{graves}. Differently from unidirectional RNNs, bidirectional models cannot be used for online speech recognition, since each prediction $\widehat{y}_{t}$ depends on the full input sequence $X$.

Nevertheless, even if the future is not accessible, we can try to roughly predict it, capturing some relevant features that help phone predictions. This principle can be implemented by means of a regularization term, as highlighted in Fig. \ref{fig:twin}. 
The idea  is to penalize forward hidden representations $\overrightarrow{h_{t}}$ that are distant from the cotemporal backward ones $\overleftarrow{h_{t}}$ ~\cite{twin_ref}. With this regard, one can add a regularization term that encourage the network to minimize the $L_{2}$ distance between forward and backward hidden states:
\begin{equation}
\Omega(\overrightarrow{\theta},\overleftarrow{\theta})=\frac{1}{N} \sum_{t=1}^N{\|\overrightarrow{h_{t}}-\overleftarrow{h_{t}}\|^2}.
\label{eq:twin}
\end{equation}
The regularization term is averaged over all time steps. In general, multiple recurrent hidden layers are stacked together to perform more robust predictions. In this case, the regularization term can be simply averaged over all the recurrent layers. 
The total objective to be minimized thus becomes a weighted sum of the NLL costs plus the regularization term:
\begin{equation*}
\widetilde{L}(X, Y; \overrightarrow{\theta},\overleftarrow{\theta}) = L(X, Y; \overrightarrow{\theta}) + L(X, Y; \overleftarrow{\theta}) + \lambda \Omega(\overrightarrow{\theta},\overleftarrow{\theta}),
\end{equation*}
where $\lambda$ is an hyper-parameter controlling the importance of the penalty term, and  $\widetilde{L}$ is the total loss that is averaged over all the sentences composing the mini-batch. 

Note that the backward states are needed only at training time, when online constraints do not arise. During testing, the part of the model computing backward states can be omitted. This leads to an architecture particularly suitable for online ASR, since it requires exactly the same amount of computations needed for standard unidirectional RNNs (at inference time). 
Another remarkable aspect of this technique is that Eq.~\ref{eq:twin} is based on the backward states $\overleftarrow{h_{t}}$, that provide a summary of the full future part of the speech sequence. This means that our method could capture not only short-term future dependencies, but also long-term ones. 





\begin{table*}[t!]
\centering
\caption{PER(\%) on the TIMIT dataset obtained with various RNN architectures and input features. Our approach \emph{UniTwin} shows stable improvement in all cases. \emph{BiDir} corresponds to the off-line recognition with bidirectional networks (reported here to provide a lower bound for the error rates of the online models).}
\label{tab:timit}
\begin{tabular}{|l|c|c|c|c|c|c|c|c|c|}
\hline
\multirow{2}{*}{} & \multicolumn{3}{c|}{MFCC}      & \multicolumn{3}{c|}{FBANK}     & \multicolumn{3}{c|}{fMLLR}     \\ \cline{2-10} 
                  & UniDir & UniTwin       & BiDir & UniDir & UniTwin       & BiDir & UniDir & UniTwin       & BiDir \\ \hline
LSTM              & 17.3   & \textbf{17.1} & 15.7  & 17.2   & \textbf{17.0} & 15.1  & 16.9   & \textbf{16.3} & 14.7  \\ \hline
GRU               & 17.9   & \textbf{17.4} & 16.0  & 18.1   & \textbf{18.0} & 15.3  & 16.9   & \textbf{16.6} & 15.3  \\ \hline
M-GRU             & 17.8   & \textbf{17.5} & 16.1  & 18.0   & \textbf{17.6} & 15.4  & 16.8   & \textbf{16.4} & 15.1  \\ \hline
Li-GRU            & 17.5   & \textbf{17.1} & 15.5  & 17.3   & \textbf{16.8} & 14.6  & 16.7   & \textbf{16.2} & 14.6 \\ \hline
\end{tabular}
\end{table*}

\subsection{Related Work}
Several methods have been proposed in the literature to approach online ASR with RNNs. A popular choice is to feed the RNN with a context window that embeds some future frames \cite{unirnn_online,peddinti_online}. Attempts have also been done to build low-latency bidirectional RNNs \cite{bidir_online1,online_zeyer,bidir_online_chen,bidir_mohamed}. The latter solutions are based on chunking the speech signal into several overlapping or non-overlapping windows. Each chunk embeds both past and future speech frames and is processed by an bidirectional RNN to perform phone predictions. These approaches, however, only account for a limited fraction of the future information, inheriting the same issues discussed for feed-forward DNNs. 

This paper proposes the use of twin regularization to improve the way online RNNs exploit the future information. Twin regularization has been recently proposed in~\cite{twin_ref}. Its effectiveness has been proved in sequence generation tasks, such as image captioning, language modeling, and monaural singing voice separation~\cite{drossos2018mad}. Some works~\cite{goyal2017z,shabanian2017variational} take a similar approach to train stochastic recurrent models with a backward running RNN. These approaches have been applied to speech synthesis, language modeling, image generation, and demonstrate that the idea of predicting future states is well-motivated, practically sound, and worth exploring.


To the best of our knowledge, this paper is the first attempt to build a predictor of future states for an online ASR model.


\section{Experimental Setup} \label{sec:setup}
In the following sub-sections, the corpora and the RNN-HMM setting adopted for the experimental activity are described.
\subsection{Corpora and Tasks}
The first set of experiments was performed with the TIMIT corpus \cite{timit}, considering the standard phoneme recognition task (aligned with the Kaldi s5 recipe \cite{kaldi}).

To validate our model in a more challenging scenario, experiments were also conducted in distant-talking conditions with the DIRHA-English dataset \cite{dirha_asru}. 
Training was based on the original WSJ-5k corpus (consisting of 7138 sentences uttered by 83 speakers) that was contaminated with a set of impulse responses measured in a real apartment \cite{rav_is16}.
The test phase was carried out with the real-part of the dataset, consisting of 409 WSJ sentences uttered in the aforementioned apartment by six native American speakers. 

Additional experiments were conducted with the CHiME~4 dataset \cite{chime3}, that is based on speech data recorded in four noisy environments (on a bus, cafe, pedestrian area, and street junction). The training set is composed of 43690 noisy WSJ sentences recorded by five microphones (arranged on a tablet) and uttered by a total of 87 speakers. 
The test set \textit{ET-real} considered in this work is based on 1320 real sentences uttered by four speakers, while the subset \textit{DT-real} has been used for hyperparameter tuning. The CHiME experiments were based on the single channel setting \cite{chime3}. 


Finally, experiments were performed with LibriSpeech~\cite{librispeech} dataset. 
We used the training subset composed of 100 hours and the \textit{dev-clean} set for the hyperparameter search. Test results are reported on the \textit{test-clean} part. 


\subsection{RNN-HMM setting}
The experiments are set up considering different acoustic features, i.e., 39 MFCCs (13 static+$\Delta$+$\Delta\Delta$), 40 log-mel filter-bank features (FBANKS), as well as 40 fMLLR features (extracted as reported in the s5 recipe of Kaldi \cite{kaldi}), that were computed using windows of 25 ms with an overlap of 10 ms.

Neural acoustic models consisted of multiple recurrent layers, that were stacked together prior to the final softmax classifier. These recurrent layers were unidirectional or bidirectional RNNs. 
Beyond standard LSTM~\cite{lstm} and GRU~\cite{gru1}, we also considered  recently proposed architectural variations~\cite{ravanelli_is17}: M-GRU~\cite{mgru} is the minimal GRU architecture based on replacing reset gate with updated gate activations, while light GRU (Li-GRU)~\cite{li_gru} directly avoids the reset gate and exploits ReLU activations for the hidden activations.

The feed-forward connections of the architecture were initialized according to the \textit{Glorot}'s scheme \cite{xavier}, while recurrent weights were initialized with orthogonal matrices \cite{orth_init}. 
Recurrent dropout was used as regularization technique \cite{drop_asru,Gal2016}. 
Batch normalization was adopted for feed-forward connections only, as proposed in \cite{laurent2016batch, ravanelli_is17}.

The optimization was done using the RMSprop algorithm running for 24 epochs. The performance on the development set was monitored after each epoch, and the learning rate was halved when the relative performance improvement went below~$0.1\%$.  
Back-propagation through time was not truncated, allowing the system to learn arbitrarily long time dependencies.

The main hyperparameters of the model (i.e., learning rate, number of hidden layers, hidden neurons per layer, dropout factor, as well as the twin regularization term $\lambda$) were optimized on the development datasets. 
In particular, we guessed some initial values according to our experience, and starting from them we performed a grid search to progressively explore better configurations. As a result, we adopted $\lambda=0.6$ for TIMIT experiments, and $\lambda=0.1$ for the other datasets. Please refer to the github repository referenced in the footnote below for more details about the considered hyperparameters.

The labels were derived by performing a GMM-based forced alignment on the original training datasets (see the standard s5 recipe of Kaldi for more details \cite{kaldi}).
During test, the posterior probabilities generated by the RNN were normalized by their prior probabilities. 
The obtained likelihoods were processed by an HMM-based decoder, that estimated the sequence of words uttered  by the speaker.

The RNN part of the system was implemented with Pytorch~\cite{paszke2017automatic}, that was  coupled with the Kaldi decoder \cite{kaldi} to form a context-dependent RNN-HMM speech recognizer.\footnote{\label{foot:code}The code is available at \url{http://github.com/mravanelli/pytorch-kaldi/}.}

\section{Results} \label{sec:exp}
In the following sub-sections, we report the experimental results obtained with TIMIT, DIRHA, CHiME, and LibriSpeech datasets. 

\subsection{Phoneme recognition on TIMIT}
To provide a thorough assessment of our methodology, several RNNs models and features are considered. Table \ref{tab:timit} shows the results obtained with TIMIT. The results with off-line bidirectional models (\textit{BiDir} columns) are reported only to provide a lower bound for the error rates that can be achieved with an online model. Moreover, to ensure a more accurate comparison, five experiments varying the initialization seeds were conducted for each RNN model and input feature. Results of Table \ref{tab:timit} are thus reported as the average \textit{phone error rates} (PER)\footnote{Standard deviations $\sigma$ range between $0.1$ and $0.2$ for all the experiments.}.

Twin regularization (\textit{UniTwin} columns) helps to improve the recognition performance, consistently outperforming standard unidirectional models (\textit{UniDir} columns) in all the considered experimental conditions. Although our method is still far from bridging the gap with off-line bidirectional models, we observe an average relative improvement of 2.1\%, which is obtained with a simple technique that does not introduce any additional computation at test time. 

Li-GRU consistently outperforms the other RNN models, as previously observed in \cite{ravanelli_is17}.  A remarkable achievement is the average PER of $14.6$\% obtained with Li-GRUs using fMLLR and fbank features. To the best of our knowledge, this result yields one of the highest published performance on the TIMIT test-set. 
 \begin{figure}[t!]
 \centering
   \includegraphics[scale=0.50]{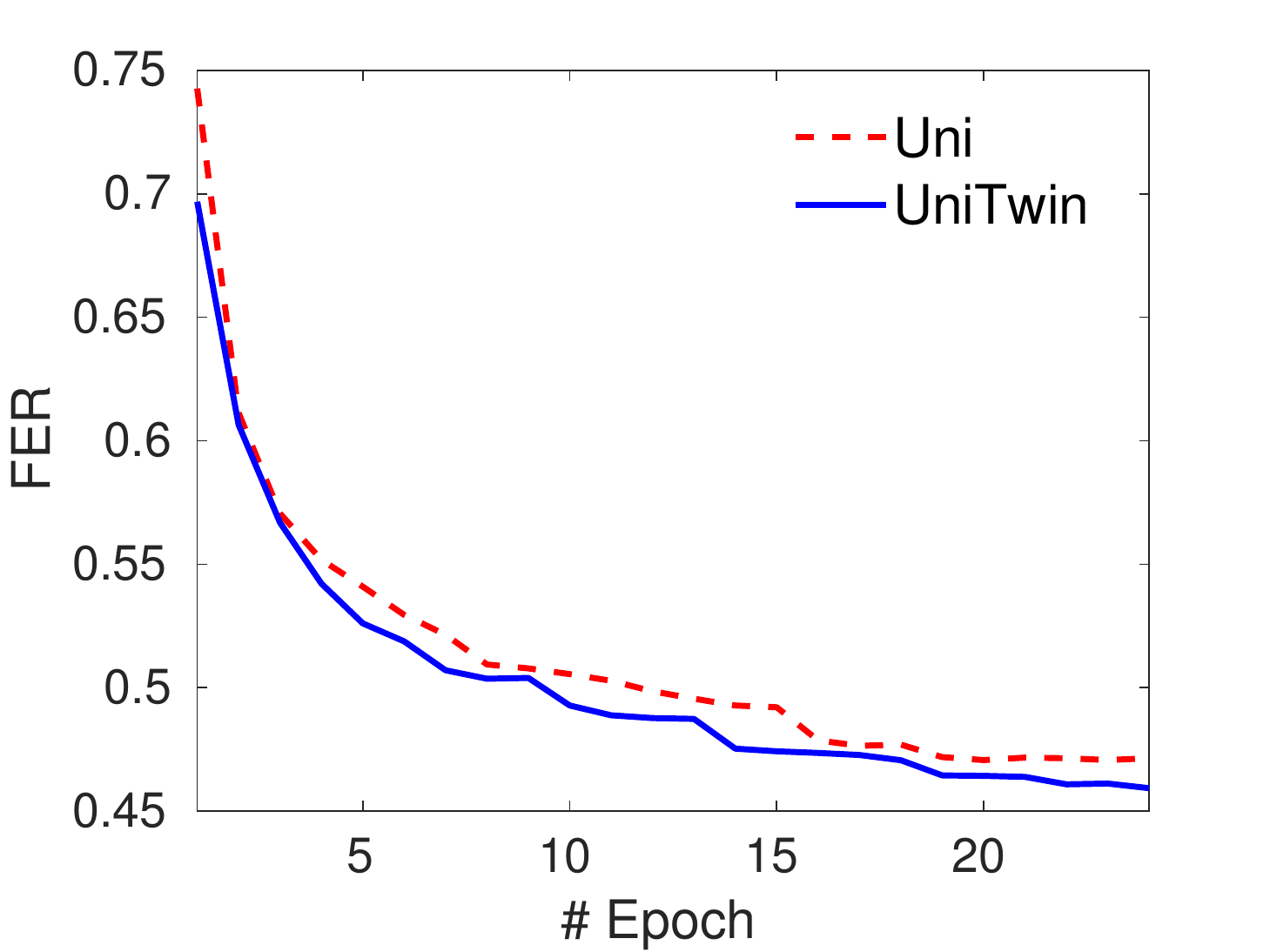}
 \caption{Learning curves for unidirectional and twin RNN models (fbank features, Li-GRU model).}\label{fig:te_cure}
 \end{figure}
We plot the learning curves for the \textit{frame-level error rates} (FER) obtained on the development set over the duration of training in Fig.~\ref{fig:te_cure}. 
Our experiments show that twin regularization converges to a better solution.

The results presented above are obtained with RNNs fed with the current frame only. Similarly to the window-based approaches described in Sec. \ref{sec:twin}, Tab. \ref{tab:cw} extends our previous results by adding a small context window that concatenates some future frames. 
\begin{table}[t]
\centering
\caption{PER(\%) on TIMIT obtained with a context windows that embeds some future frames (Li-GRU model, fbank feats).}
\begin{tabular}{|l|c|c|}
\hline
\# Future Frames         & UniDir & UniTwin \\ \hline
0 frames       &    17.3     &     \textbf{16.8 }  \\ \hline
5 frames &    16.5    &   \textbf{16.1 }     \\ \hline
10 frames  &    16.8    &    \textbf{16.5}     \\ \hline
15 frames &    17.5    &     \textbf{16.9}    \\ \hline
\end{tabular}
\label{tab:cw}
\end{table}
The table shows that a small look-ahead context window embedding 5 or 10 future frames is helpful to improve the ASR performance. Interestingly, our method outperforms standard unidirectional RNNs even under this experimental condition. This achievement confirms that twin regularization can focus on long-term future dependencies, providing useful information also when some look-ahead frames embed a short-term future context. This allows our method to be used in conjunction with previous window-based approaches.

\subsection{Word recognition on other datasets}
As a last experiment, we extend our previous achievements to more realistic ASR tasks.
To test our technique into a complex acoustic scenario, Tab. \ref{tab:dirha} reports the \emph{word error rate} (WER) obtained with the DIRHA dataset. For the sake of compactness, only the results with MFCCs are reported. It is worth mentioning that we obtained a similar experimental evidence using fbank and fMLLR features. 

\begin{table}[h]
\centering
\caption{WER(\%) for the DIRHA dataset (MFCC feats).}
\label{tab:dirha}
\begin{tabular}{|l|c|c|c|}
\hline
            & UniDir & UniTwin & BiDir \\ \hline
LSTM       &   32.9     &   \textbf{32.5}      &  27.8     \\ \hline
GRU &    30.2    &    \textbf{29.6}     &   27.2    \\ \hline
Li-GRU &  29.2      &    \textbf{28.7}     &  26.9    \\ \hline
\end{tabular}
\end{table}

\begin{table}[h]
\centering
\caption{WER(\%) for CHiME (ET-Real) and Librispeech (Test-Clean) using Li-GRU models and MFCC feats.}
\label{tab:others}
\begin{tabular}{|l|c|c|c|}
\hline
            & UniDir & UniTwin & BiDir \\ \hline
CHiME       &  23.7      &   \textbf{23.0}      &  19.2     \\ \hline
LibriSpeech &   10.4     &   \textbf{10.2}      &  9.2     \\ \hline
\end{tabular}

\end{table}

We conclude from this experiment that twin regularization is also effective in challenging acoustic conditions characterized by the presence of both noise and reverberation. 

To further test its robustness in noisy environments, we also performed some experiments with the CHiME dataset (see first row of Tab. \ref{tab:others}). Moreover, to provide evidence on a larger vocabulary task, the second row of Tab. \ref{tab:others} reports the results achieved with LibriSpeech. These results are obtained with the 100 hours subset decoded with the \textit{tgsmall} language model (see Kaldi s5 recipe \cite{kaldi}). 
Our experiments target the online ASR scenario, therefore the results reported in the table do not consider complex techniques as multi-microphone processing, data-augmentation. Neither system combination nor lattice rescoring are used here. 
It is however worth noting that the effectiveness of the proposed approach is one more time confirmed.  

\section{Conclusions} \label{sec:conc}
This paper explored the use of twin regularization for predicting the future states of an online RNN-HMM speech recognition. The proposed technique, that encourages forward hidden representations to be predictive of the future, has shown to be effective in several experimental conditions.

An average relative performance improvement of 2\% is obtained over a standard unidirectional RNNs. The improvement is consistent across datasets, architectures, and input features. 
Furthermore, our proposed technique is simple and does not add any additional computational cost at test time.

A noteworthy aspect of our method is that it also accounts for long-term future dependencies, which differs from current dominant approaches based on short-term context windows. This offers the possibility of using twin regularization in conjunction with existing techniques. 



\section{Acknowledgment} \label{sec:conc}
We would like to thank Titouan Parcollet (Universit\'e d'Avignon), Kyle Kastner (Universit\'e de Montr\'eal) and Maurizio Omologo (Fondazione Bruno Kessler) for their helpful comments. This research was enabled in part by support provided by Calcul Qu\'ebec and Compute Canada (www.computecanada.ca).
\vfill \pagebreak
\bibliographystyle{IEEEtran}

\bibliography{mybibfile}

\end{document}